\documentclass[useAMS,usenatbib]{mnras}
\usepackage{multirow}
\usepackage{hhline}
\usepackage{times}
\usepackage{graphics,epsf}
\usepackage[T1]{fontenc}
\usepackage{aecompl} 
\usepackage{amsmath}               
\usepackage{placeins}

\usepackage{amsfonts}               
\usepackage{amssymb}                
\usepackage{epsfig}                 
\usepackage{epstopdf}
\usepackage{rotating}
\usepackage{color}
\usepackage{tabularx}
\usepackage{url}
   
\def \apj{ApJ}
\def \aap{A\&A}
\def \aj{AJ}

\def \mnras{MNRAS}
\def \apjl{ApJ Lett.}
\def \apjs{ApJ Suppl. Ser.}
\def \nat{Nature}

\def \araa{ARA\&A}

\def \icarus{Icarus}

\begin{document}

\title{Generalized Hill-Stability Criteria for Hierarchical Three-Body Systems
at Arbitrary Inclinations }

\author[Grishin et. al.]{Evgeni  Grishin, Hagai B. Perets, Yossef Zenati \& Erez Michaely
\\Physics department, Technion - Israel institute of Technology, Haifa,
Israel 3200002
\\eugeneg@campus.technion.ac.il; hperets@physics.technion.ac.il}
\maketitle

\begin{abstract}
A fundamental aspect of the three-body problem is its stability. Most stability studies have focused on the co-planar three-body problem, deriving analytic criteria for the dynamical stability of such pro/retrograde systems. Numerical studies of inclined systems phenomenologically mapped their stability regions, but neither complement it by theoretical framework, nor provided satisfactory fit for their dependence on mutual inclinations. Here we present a novel approach to study the stability of hierarchical three-body systems at arbitrary inclinations, which accounts not only for the instantaneous stability of such systems, but also for the secular stability and evolution through Lidov-Kozai cycles and evection. We generalize the Hill-stability criteria to arbitrarily inclined triple systems, explain the existence of quasi-stable regimes and characterize the inclination dependence of their stability. We complement the analytic treatment with an extensive numerical study, to test our analytic results. We find excellent correspondence up to high inclinations $(\sim120^{\circ}$), beyond which the agreement is marginal. At such high inclinations the stability radius is larger, the ratio between the outer and inner periods  becomes comparable, and our secular averaging approach is no longer strictly valid. We therefore combine our analytic results with polynomial fits to the numerical results to obtain a generalized stability formula for triple systems at arbitrary inclinations. Besides providing a generalized secular-based physical explanation for the stability of non co-planar systems, our results have direct implications for any triple systems, and in particular binary planets and moon/satellite systems; we briefly discuss the latter as a test case for our models.

\end{abstract}
\begin{keywords}
celestial mechanics~\---~minor planets, asteroids, general~\---~planets and satellites: dynamical evolution and stability
\end{keywords}

\maketitle

\section{INTRODUCTION}

The three-body problem is an old topic in celestial mechanics, with
wide astrophysical applications in the Solar system and beyond \citep{V08, I97, Holman97}. Hierarchical triple systems are systems in which an inner binary orbits a more
distant object on an outer orbit, with some mutual given inclination
between the inner and outer orbits. A fundamental aspect of a hierarchical
triple system is its stability. A system is considered stable if no
collision or escape of one of the bodies occurs after a large number
of orbital periods. The natural length scale of stability is the mutual Hill
radius $r_{H}=a_{out}(\mu/3)^{1/3}$, where $a_{out}$ is the distance
from the distant outer body with mass $m_{out}$ to the center of
mass of the binary and $\mu\equiv(m_{1}+m_{2})/m_{out}$ is the binary
to perturber mass ratio. The Hill radius is centered on the center of mass of the inner binary. In the case of the test particle limit $m_1 \gg m_2$ the Hill radius is centered on $m_1$.  If the inner binary semi-major axis (SMA) is larger than
the Hill radius $a_{in}>r_{H}$, the binary is unstable, since tidal
forces from the perturber shear apart the binary. Conversely, if $a_{in}\ll r_{H}$,
then the binary is stable and the perturbations from $m_{\text{out}}$ are
small. Early studies of \emph{co-planar} triple systems 
\citep{1967torp.book.....S,Henon70, HB91} have
shown that prograde orbits are stable for $a_{in}\approx0.5r_{H}$
while retrograde orbits are stable for twice the distance.

\begin{figure}
\begin{centering}
\includegraphics[width=0.45\textwidth]{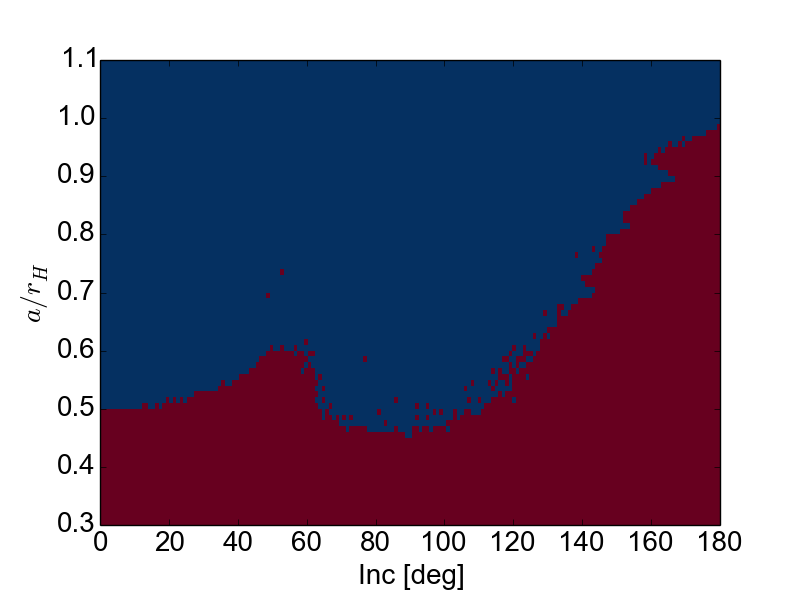}

\caption{Numerical 
stability map. The initial conditions for stability map are described in sec. \ref{subsec:Numerical-set-up}. Left: Blue pixel indicates stable orbit, red pixel \label{fig:1}
is unstable orbit. The stability map is morphologically similar to \protect\cite{HB91} (their Fig. 15).}
\end{centering}
\end{figure}

The first analytic study to explore the critical stability radius
at \emph{arbitrary} inclination was done by Innanen \citep{1979AJ.....84..960I,I80}. He finds that the critical stability radius
is an \emph{increasing} function of the inclination, thus 
the most stable orbits are retrograde, consistent with previous
results of co-planar  orbits.
However, numerical simulations show that the critical radius is compatible
with the analytical expectation only for moderately inclined orbits.
For highly inclined orbits, the critical radius starts to decrease at $\sim60^{\circ}$,
and increases again only at higher inclinations, forming form a bowl-like
shape (see Fig. \ref{fig:1} and Fig. 15 of \citealp{HB91}).

The general equations of motion of the three-body problem cannot be
solved analytically \citep{ValtonenBook2006}, however secular averaging
analysis can be used to describe a wide range of cases. For hierarchical
systems, pioneering works of \citet{Lidov62} and \citet{Kozai62}
have shown that the torque of the outer binary can induce significant
quasi-periodic oscillations in the inclination and eccentricity of
the inner binary over long, secular timescales, under certain configurations. 

The Lidov-Kozai (LK) mechanism is obtained by double averaging (DA)
over the orbits. The averaging process does not take into account
short period terms that affect the secular evolution. \citet{CB04} considered corrections due to evection and Lunar theory
for circular perturbers in the context of irregular moons around giant
planets. Recently, a critical analysis of the DA technique and its
correction, as well as generalization of \cite{CB04} for arbitrary inclination
and eccentricity was studied by \cite{Katz16}.

In previous analytic studies of Hill stability, the impact of secular
evolution, and in particular Lidov-Kozai cycles were not fully addressed.
Numerical studies provided insights and phenomenological mapping for
the stability criteria of such systems \citep{HB91,Nesvorny03,Frouard10,Domingos06},
but did not explain their physical origin, nor provided analytic derivations
and satisfactory fits for the the dependence of stability on the mutual
inclinations. In this paper we account for secular evolution in exploring
the stability of triple systems (namely the LK mechanism and evection),
and provide a physical understanding of its behaviour. In addition,
we perform an extensive numerical three-body study of the stability
criteria, confirming our analytic solutions (finding excellent correspondence
up to high inclinations) and complement them in regimes where the
secular averaging approach we use is no-longer valid. 

For hierarchical two-planet systems, the Lagrange-stability  using resonance overlap has been studied in \citet{Veras04, M08}, while empirical stability criteria are obtained in \citet{Ma2001} and more recently in \citet{P15}. In this paper we deal with the Hill-stability of  hierarchical systems in mass.

The paper is organized as follows. In Sec. \ref{sec:2}
we briefly provide the background for previous analytic study of the
instantaneous Hill stability at arbitrary inclinations  based on \cite{I80}. In Sec. \ref{sec:Novel-Lidov-Kozai-Hill-secular} We first describe the LK mechanism, and then couple it to stability
analysis. We then show how secular evolution
affects the results, explaining the discrepancy between
\cite{I80} and \cite{HB91}.
 In Sec. \ref{sec:Evection-and-Lunar} we explore how 
evection further improves the stability analysis. In Sec.
\ref{sec:Numerical-parameter-space}, we numerically integrate the three-body problem to confirm and complement our analytic approach, providing a generally
useful fitting formula for triple stability at arbitrary inclinations.
Finally we discuss our results and their implications and summarize
in Sec. \ref{sec:Discussion-and-summary}. 

\section{ \emph{Instantaneous} Hill
stability criteria at arbitrary inclinations: Coriolis Asymmetry }
\label{sec:2}
Before deriving a generalized stability criteria which accounts for
\emph{secular} processes, let us first review the main results of
\citet{I80}'s work on the Coriolis asymmetry, and derive an expression
for the \emph{instantaneous} stability criteria in a modern fashion.
This expression, as remarked above, fails to correctly capture the
observed stability behaviour, however it is not incorrect by itself,
but rather has to be coupled to secular effects, as we show in the
next sections. 

Consider the restricted, mass hierarchical three-body system. We will
use "star" of mass $m_{out}$, "planet" of mass $m_{1}$ and
"satellite" of mass $m_{2}$ for the hierarchical masses $m_{out}\gg m_{1}\gg m_{2}$,
but alternatively, it could apply to other astrophysical systems.
The planet revolves around the star at SMA $a_{out},$
and the satellite revolves around the planet at SMA $a_{in}\ll a_{out}.$
For convenience, we omit the subscript "in" for the orbital elements of the inner binary (e.g. the assumption
on hierarchical system is $a\ll a_{out}$).

Consider a binary with arbitrary inclination. The instantaneous distances
of the satellite from the star and the planet are $\boldsymbol{R}$
and $\boldsymbol{r}$ respectively. The acceleration of the
satellite in the reference frame centered on the star is given by \citet{I80}

\begin{equation}
\boldsymbol{a}=\ddot{\boldsymbol{R}}+\boldsymbol{\dot{\Omega}}_{out}\times\boldsymbol{r}+\boldsymbol{\Omega}_{out}\times(\boldsymbol{\Omega}_{out}\times\boldsymbol{r})+2\boldsymbol{\Omega}_{out}\times\boldsymbol{v}_{r},\label{eq:acceleration1}
\end{equation}
 where $\boldsymbol{\Omega}_{out}$ is the angular frequency vector
around the star, $\boldsymbol{v}_{r}=\omega_{1}\boldsymbol{r}$ is
the velocity of the satellite and $\omega_{1}$ is the angular frequency
around the planet\footnote{The term 'mean motion' is also frequently used, and usually denoted $n_1$ for the planet and $n_{out}$ for the star.}. The first term is $\ddot{\boldsymbol{R}}\boldsymbol{=\Omega}_{out}\times(\boldsymbol{\Omega}_{out}\times\boldsymbol{R})$,
the gravitational acceleration from the star, the second term is the
change of angular frequency, the third term is the gravitational acceleration
from the planet, and the last term is $2\boldsymbol{\Omega}_{out}\omega_{1}\boldsymbol{r}\cos i$,
the Coriolis term, where $i$ is the inclination angle of the binary.
For constant $\boldsymbol{\Omega}_{out}$, the second term vanishes.
At the critical radius, tidal forces from the Sun and other forces
equal the gravitational pull of the planet. Thus, at the critical
radius, the acceleration is zero, and Eq. (\ref{eq:acceleration1})
is an algebraic equation for the minimal angular frequency (or maximal
radius) 
\begin{equation}
\omega_{1}^{2}-2\omega_{1}\Omega_{out}\cos i-3\Omega_{out}^{2}=0\label{eq:rcrit1}
\end{equation}
For prograde orbit ($i=0$) , $\omega_{1}=3\boldsymbol{\Omega}_{out}$
and the critical radius is $r_{c}=r_{H}/3^{1/3}\approx0.7r_{H}.$
For retrograde orbit ($i=180^{\circ}$), $\omega_{1}=\boldsymbol{\Omega}_{out}$
and the critical radius is $r_{c}=3^{1/3}r_{H}=1.44r_{H}$. Thus,
the Coriolis force is a stabilizing force for retrograde orbits and
destabilizing force for prograde orbits.

It is possible to find the limiting radius for any inclination. \cite{I80}
finds the ratio of the prograde to retrograde orbit for each inclination.
Alternatively, we solve the quadratic equation (\ref{eq:rcrit1})
and find the limiting radii explicitly. The only positive solution
to Eq. (\ref{eq:rcrit1}) is $\omega_{1}(i)/\Omega_{out}=g(i),$
where $g(i)\equiv\cos i+(3+\cos^{2}i)^{1/2}$. Note that the Coriolis
asymmetry is evident in the first term, and that the solution reduces
to $\omega_{1}=3\Omega_{out}$ for $i=0$ and $\omega_{1}=\Omega_{out}$
for $i=180^{\circ}.$ The limiting radius is then 
\begin{equation}
r_{c}(i)=\left(\frac{Gm_{1}}{\omega_{1}^{2}(i)}\right)^{1/3}=3^{1/3}r_{H}g(i)^{-2/3}\label{eq:rcrit2}
\end{equation}
Note that $g(i)$ in a monotonic decreasing function for $i\in[0,\pi]$,
thus $r_{c}(i)$ is an increasing function. 

The limiting radii derived in (\ref{eq:rcrit2}) is obtained by taking
the acceleration of the satellite to be zero. In reality, the binary
could break apart before it reaches the limiting radius $r_{c}$.
Indeed, numerical integrations (e.g. \citealp{HB91,Nesvorny03,CM90}) show that the transition from stable to unstable binary is $\approx0.5r_{H}$
for prograde orbits, and $\approx1r_{H}$ for retrograde orbits. Thus,
a fudge factor $q\le1$ can be introduced to effectively rescale the
stability limit of Eq. (\ref{eq:rcrit2}). Note that for prograde orbit the
stability limit is consistent with the critical Jacobi integral that
allows escape \citet{CM90}. 

In the latter discussion, one has to remember that the three body
problem is chaotic, and the transition boundary from stable to unstable
orbit is not strictly a line, but rather has a complex, fractal, structure.
The final fate of the orbits near the stability limit sensitively
depends on the initial conditions, and large number of orbits have
to be integrated in order to get a statistical understanding of the
stability limit. We explore the parameter space numerically in Sec.
\ref{sec:Numerical-parameter-space}.

The numerical integrations of \cite{HB91} fit the expectations for almost
co-planar inclinations (i.e. such that $\sin i$ is small). However,
for high inclinations (i.e. such that $\sin i$ is large), the critical
radius significantly deviates from the analytical expectations. In
particular, the critical radius is \emph{decreasing} for inclinations
between $\sim60^{\circ}$ and $\sim90^{\circ}$, and is substantially
lower than expected for inclinations up to $\sim150^{\circ}$(see
Fig. \ref{fig:1}).

\section{Novel Lidov-Kozai-Hill
\emph{secular} instability criteria at arbitrary inclinations }
\label{sec:Novel-Lidov-Kozai-Hill-secular}

In order to derive a generalized stability criteria which accounts
for \emph{secular} processes, we first review the secular LK
mechanism. We then  couple the  stability criteria given in Eq. (\ref{eq:rcrit2}) to the secular LK mechanism to modify and derive  a \emph{generalized} \emph{secular}
stability criteria and compare it with a numerically integrated stability
map. As we show, the novel Lidov-Kozai-Hill stability criteria qualitatively
captures the instability properties at large inclinations. We further
correct this criteria by accounting for evection (Lunar theory) processes
in the next section. 

\subsection{Lidov-Kozai mechanism }

The presence of a third, distant body in hierarchical system causes
the orbital elements of the binary to change with time. The angles
of ascending node $\Omega$ and argument of periapse $\omega$ precess,
while other orbital elements oscillate quasi-periodically with small
magnitude, that depends of the ratio $a/a_{out}.$ On average, the
energy and angular momentum of the binary are conserved over orbital
period of the distant body.

In the Lidov-Kozai (LK) resonance, the argument of periastron $\omega$
librates, rather than oscillates. Even for orbits not in resonance,
the LK mechanism exchanges the binary eccentricity and inclination,
obtained for large enough initial inclination $i$. For hierarchical
triple systems the typical  LK cycle timescale is 
\begin{equation}
T_{LK}\approx\frac{m_{out}+m_{1}+m_{2}}{m_{out}}(1-e_{out}^{2})^{3/2}\frac{P_{out}^{2}}{P}\label{eq:tlk1}
\end{equation}

For the restricted three body problem and $e_{out}=0$, it reduces
to 
\begin{equation}
T_{LK}\approx\sqrt{3}P_{out}\left(\frac{a}{r_{H}}\right)^{-3/2}=3P\left(\frac{a}{r_{H}}\right)^{-3}\label{eq:tlk2}
\end{equation}
For most cases, the LK timescale is a \emph{secular} timescale, much
larger that the orbital period. Thus, the DA method
is satisfactory. However, in our case, $a$ is relatively comparable
to $r_{H}$, thus, the LK timescale is not much larger than the binary
period $P$, and consequently the system is only marginally (or quasi)
secular. We discuss the implications and caveats of the quasi secular
regime in Sec. \ref{sec:Discussion-and-summary}

In the quadrupole DA approximation, the minimal inclination required
for the LK mechanism is $i_{c}=\arcsin\sqrt{2/5}$, or $\approx39.2^{\circ}$
for prograde orbits and $140.8^{\circ}$ for retrograde orbits. In addition, the $z$ component angular momentum
of the binary, $L_{z}=\sqrt{1-e^{2}}\cos i$ is constant. Thus, the
maximal eccentricity attained during a KL cycle is
\begin{equation}
e_{max}(i,i_{c})=\sqrt{1-\frac{\cos^{2}i}{\cos^{2}i_{c}}}=\sqrt{1-\frac{5}{3}\cos^{2}i}\label{eq:emax}
\end{equation}
The larger the initial inclination $i$ corresponding to initial eccentricity $e=0$ is, the larger is $e_{max}$.
In addition, $e_{max}\to1$ as the inclination approaches to $90^{\circ}$. 

Using further octupole expansion beyond the quadrupole approximation, it was shown that under specific conditions even higher inner eccentricities
can be obtained \cite{2001Icar..150..303F}, and the inner orbit
inclination can flip its orientation from prograde, to retrograde
\cite{EKL-naoz2011} (see \cite{Naoz2016review} for a
recent review and references therein of this eccentric LK
(EKL) regime). Moreover, the secular averaging method itself fails
at some point, and the systems can behave more erratically, leading
to large eccentricity changes on dynamical timescales, and even more
extreme eccentricities can be obtained \cite{2012ApJ...757...27A}.
Here we only take into account the quadrupole term, other regimes
are beyond the scope of this manuscript and will be discussed elsewhere.

\subsection{Lidov-Kozai-Hill stability}

\begin{figure*}
\begin{centering}
\includegraphics[width=0.45\textwidth]{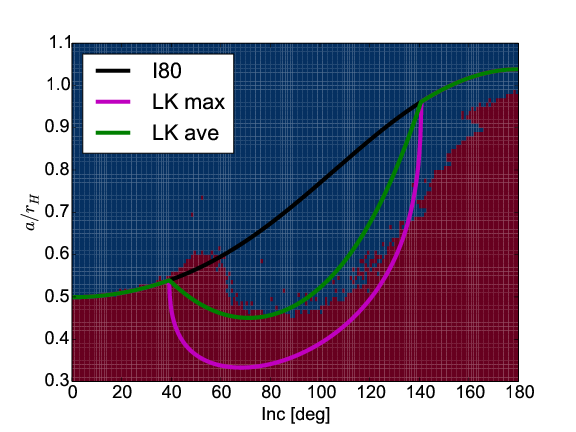}\includegraphics[width=0.45\textwidth]{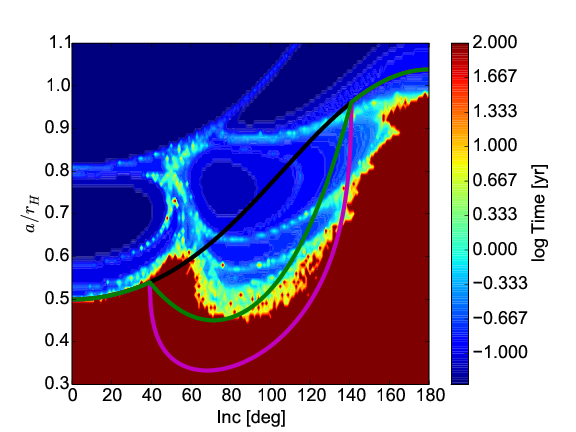}
\par\end{centering}
\caption{ \label{fig:2}Left: Same as in Fig \ref{fig:1}. The black line is the original curve given in Eq. (\ref{eq:rcrit2}) (normalized to $0.5r_{H}$ at zero inclination).
The magenta line is the modified curve due to LK mechanism with $f_{max}$. The green line is is the modified curve due to LK mechanism
with $f_{ave}(e)=1/(1+0.5e^{2})$. Right: Break up times for the stability
map. The stable region has time $=100$yr.}
\end{figure*}

Previous numerical simulations indicate that while the Coriolis asymmetry
describes well co planar orbits, other effects besides the Coriolis
asymmetry must be taken in to account for orbits with high inclination. 

The original derivation of \cite{I80} is for circular orbits. In reality,
secular perturbations due to the LK mechanism drives the binary to
high eccentricities, that affect the limiting radii of stability.
Given a maximal eccentricity $e_{max}$, the limiting radii modified
by the LK mechanism is 
\begin{equation}
r_{c}^{(LK)}(i,e_{max})\equiv r_{c}(i)\cdot f(e_{max})\label{eq:rcrit3}
\end{equation}
Where $r_{c}(i)$ is given by Eqn. (\ref{eq:rcrit2}), $f(e_{max})$
is some function of $e_{max},$ such that $f(0)=1.$ For secular timescales, the critical radius can be satisfied at the apoapsis, or the maximal
distance of the orbit, i.e. if $r_{c}^{(LK)}(i)(1+e_{max})=r_{c}(i)$,
thus $r_{c}^{(LK)}=r_{c}(i)/(1+e_{max})$, or $f=f_{max}(e_{max})=1/(1+e_{max})$.

\citet{Carruba02} and \citet{Per+09}  pointed out the importance of
the LK secular evolution for the stability. Here we follow-up on this
issue and consider the effects of secular evolution, couple them to
the \cite{I80} approach and derive a generalized criteria which well captures
the results from three-body simulations. In order to explore this analytically
we first repeat the numerical experiment of \cite{HB91} with a different integrator
and more refined resolution (as described in sec. \ref{subsec:Numerical-set-up}).
We obtain a highly resolved stability map, to be compared with the
analytic results we derive. 

The left panel of Fig. \ref{fig:2} shows
this stability map. Red pixel indicates stable orbit for $100$ yr,
while blue pixel is unstable orbit. Generally, the boundary of stability
is chaotic, but the chaotic regions are narrow and still a boundary
line could be traced. The boundary increases with inclination, up
to $\sim60^{\circ}$, then it abruptly decreases, and start to slowly
increase again for retrograde orbits, up to $\sim1r_{H}$ for inclinations
near $180^{\circ}$. The black curve is
given in Eq. \ref{eq:rcrit2}; it is increasing with increasing inclination,
contrary to the boundary of the stability. The magenta curve is the
correction due to LK mechanism, $r_{c}^{(LK)}$ with $f_{max}=(1+e_{max})^{-1}$
in Eq. (\ref{eq:rcrit3}) (i.e. ``LK max''). The curve is bowl-shaped
in the LK dominated region of high inclination, consistent with decrease,
and subsequent increase of the stability boundary. We note that even
though we have used a different integrating scheme and different masses
and integration times, the results of the stability map reproduce
the results of HB91. 

The right panel of Fig. \ref{fig:2} shows the break up time of the
same simulated map. Dark red times are stable orbits with $t=100$yr.
The shortest break up times are blue (comparable to dynamical times),
while longer break up timescales are more red. We see that generally
in the region where $i\ge60^{\circ}$the break up times are longer,
indicating that the stability limit in this region is dominated by
(quasi) secular effects. The additional structures of the break up
timescale in the unstable region are probably due to additional resonances
of the system, are out of the scope of this paper and will be studied
elsewhere.

Though morphologically similar to the simulated stability boundary,
the LK max curve has discrepancies both in the ``turnoff point''
where the LK mechanism is effective, and in the depth of the bowl. 

A possible explanation to the discrepancy of the depth of the bowl
is that near the stability limit, the timescales involved are not
secular, but rather comparable to the dynamical timescale, i.e. the
binary period $P$. In this case, the effective radius is not the
apoapsis, but rather the average separation $\langle a\rangle_{P}=a(1+0.5e^{2})$ \citep{Frouard10}. The maximal average distance is
found by setting $e=e_{max},$ thus the correction factor for taking
the average distance is $f_{ave}(e_{max})=1/(1+0.5e_{max}^{2})$.
The green curve in Fig. \ref{fig:2} shows the correction due to LK
mechanism, $r_{c}^{(LK)}$ with $f_{ave}$ in Eq. (\ref{eq:rcrit3})
(i.e. "LK ave"). It maintains the morphology of the shape of the
stability boundary, though still is an effective shift in the critical
inclinations of the LK mechanism.

We ran the same stability map for longer times of $10^{4}$yr, $100$
times more that the previous integration time, and $\sim10^{3}$ times
more than the typical LK time-scale. Generally, the morphology of
the stability map remains the same. The most notable difference is
that some of the orbits in the chaotic region that were stable after
$100$ years are unstable after $10^{4}$ years. Besides this (expected)
difference, the picture remains the same. We conclude that the stability
grid is (almost everywhere) long lived, and assume hereafter that
the same picture remains unaltered even for longer integration times.
We conclude that the end time of $100$ yrs fully captures the
picture of the stability map, and almost no information is lost with
the integration time of $100$ years (e.g., the "bump" near $60^{\circ}$
is real and not a numerical artifact of short integration times).

\section{Evection and Lunar theory corrections}
\label{sec:Evection-and-Lunar}
Taking into account the secular LK mechanism successfully improved
the stability limit of inclined orbits with respect to numerical integrations,
both in scale and shape. However, the discrepancy of the effective
"turn off" point where the LK mechanism is effective remains.
The critical inclination where the slope of the stability limit changes
is near $\sim60^{\circ}$ in the numerical stability map, while in
the LK mechanism, the critical inclination angle is much lower ($i_{c}\approx39.2^{\circ}).$

Interestingly, the inclination distribution of prograde irregular
satellites in the Solar System has a cut-off near $60^{\circ}$. In addition, the entire range of inclinations of $60^{\circ}<i<140^{\circ}$
is devoid of satellites \citep{Carruba02,2005AJ....129..518S}. In
order to tackle the problem, we consider additional perturbations
to the system, namely the evection resonance (ER) and Lunar theory
(LT). 

ER is a semi-periodic, rather than secular, perturbation to a satellite's
orbit with a resonant angle $\lambda_{out}-\varpi$ where $\lambda_{out}$ is
the mean longitude of the planet and $\varpi = \omega + \Omega$ is the longitude of pericenter. The
mean longitude is a fast angle that varies with the orbital time-scale.
The averaging over the orbital period does not take into account the
ER \citep{CB04}. 

\begin{figure*}
\includegraphics[width=0.45\textwidth]{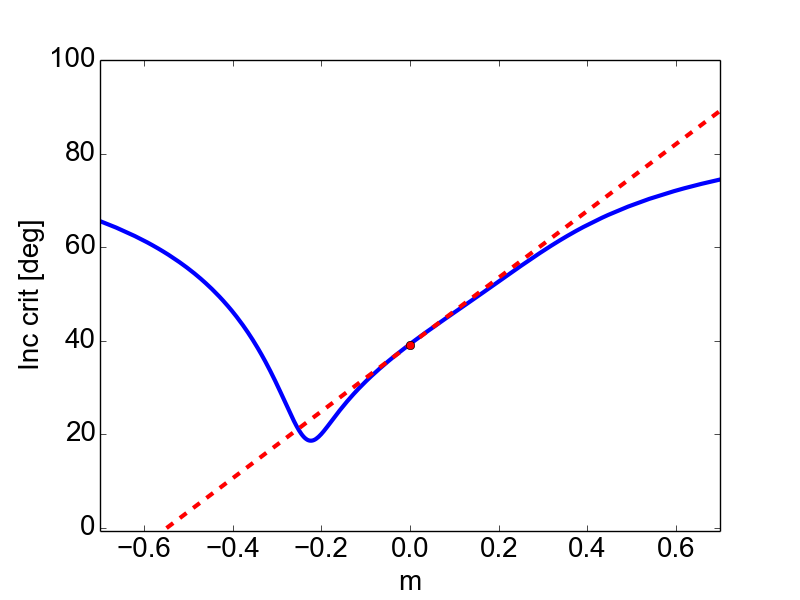}
\includegraphics[width=0.45\textwidth]{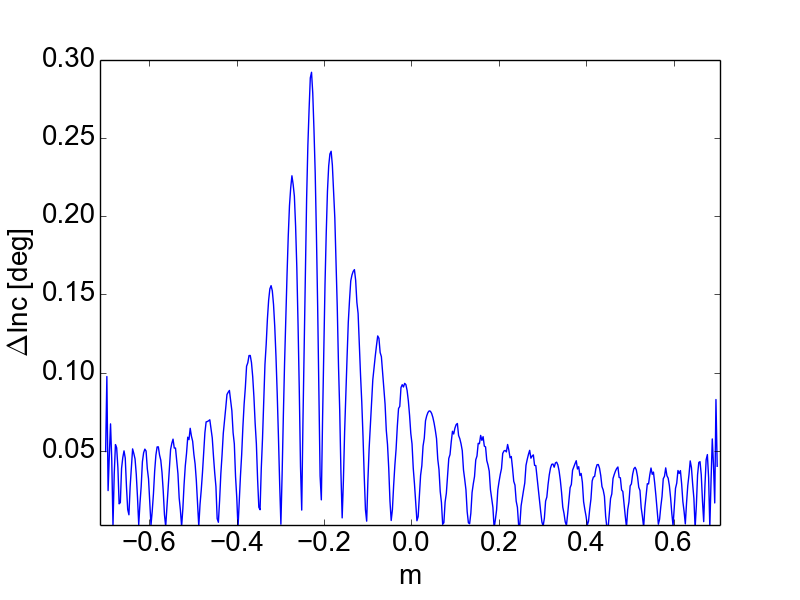}\caption{\label{fig:Left:i_crit_of_m}Left: Numerical solution of Eq. (\ref{eq:icrit})
(blue). First order analytical solution of of Eq. (\ref{eq:icrit})
(red, dashed). The red dot is the classical LK critical inclination.
Right: Deviation of the numerical solution to the polynomial least
squares fit of order $40$. The RMS value of the polynomial fit is
$\Delta I_{RMS}=0.08^{\circ}$.}
\end{figure*}

One of the consequences of ER is a change in the  precession of the longitude of pericenter $\varpi$. \footnote{Note that $\omega$ is the argument of periapse of the satellite-planet
system, not to be confused with the angular frequency of the satellite
$\omega_{1}$. Similarly, $\Omega$ is the line of nodes of the satellite-planet
system, not to be confused with the angular frequency of the planet
around the star $\Omega_{out}$.} From secular theory, the precession rate $\dot{\varpi}=(3/4)\Omega_{out}^{2}/\omega_{1}$
has an error of $\sim50\%$ from its observed value for the Earth-Moon system. Newton considered
this discrepancy as a major flaw in his theory of gravitation. Later
it has been realized that the large errors arise from evection terms
that remain even after the averaging of the orbital motion 
\citep{CB04,Katz16}. The strength of the evection is determined
by a parameter 
\begin{equation}
m=\Omega_{out}/\omega_{1}\label{eq:m}
\end{equation}
For the moon, it it about $\sim1/13$.

Various works on LT \citep{Brouwer61book, SahaTremaine93, Fr13}
have derived the precession of the pericenter in a form of power series
in $m$
\begin{equation}
\dot{\varpi}=\Omega_{out}\left(\frac{3}{4}m+\frac{225}{32}m^{2}+\frac{4071}{128}m^{3}+\cdots\right)\label{eq:precession1}
\end{equation}
The second term arises from ER, while higher order terms are calculated
from LT \citep{Tisserand1894}. For negative $m$,
the orbit is retrograde and $\varpi$ slowly precesses or even librates. 

Higher order averaging methods are available to correct the slow frequencies (e.g. \citealp{Beauge06}). Here we use the empirical, but nevertheless accurate method of \cite{CB04}. Using a development of the disturbing function to quadrupole order,
\cite{CB04} have shown that the additional terms cause additional precession
of the orbital elements.  One  important consequence of the ER
and LT terms is the effective shift of the critical inclination for
LK resonance. The equation for the precession of apsides is Eq. (38)
of \cite{CB04} 
\begin{eqnarray}
\frac{d\omega}{d\tau} & = & h_{0}(i)+mh_{1}(i)\cos i+\sum_{k=2}^{10}C_{k+1}m^{k}\cos^{k}i\label{eq:icrit}
\end{eqnarray}
where $h_{0}(i)={(3/4)}(2-5\sin^{2}i)$, $h_{1}(i)=(225/64)(2-\sin^{2}i)+(9/32)\left[3\sin^{2}i-\cos(2i)\right]$,
$\tau\equiv(1-e_{out}^{2})^{-3/2}m\Omega_{out}t$ is the dimensionless
time. Various values of the   coefficients $C_{k+1}$ are given in the literature
\footnote{The earliest list is in p. 233 of \cite{Tisserand1894}. \cite{SahaTremaine93}
use slightly different coefficients. \cite{1971AJ.....76..273D}
compares various coefficients derived in the late 19th and early 20th
century, namely Tisserand and Hill (see table III). \cite{Fr13}
is a more modern work. }. For clarity, we write down and discuss our choice of the coefficients in appendix \ref{sec: app A}.

LK resonance occurs whens $d\omega/d\tau=0$, which determines the
critical inclination. The $h_{0}(i)$ term is responsible for the
classical LK resonance, occurs at $i_{c}=\arcsin\sqrt{2/5}\approx39.2^{\circ}$.
The $h_{1}(i)$ terms comes from the ER terms in the disturbing function.
Higher order terms $C_{k}$ come from LT. 

Eq. (\ref{eq:icrit}) has been solved iteratively in \cite{CB04}. However,
for $m$ larger than $\sim0.25$ the iterative method diverges. Thus,
we solve Eq. (\ref{eq:icrit}) numerically by substituting each pair
$(i_{j},m_{k})$ in $A_{jk}\equiv|d\omega/d\tau(i_{j},m_{k})|$,
given in Eq. (\ref{eq:icrit}) and seeking the values of $(i_{j},m_{k})$
for which $A_{jk}<\epsilon$ where $\epsilon$ is some small number.
We took $\epsilon=0.005$ such that each $m$ has a solution $m(i)$. Both methods are consistent for small $m$'s and produce the same results where the iterative method of \cite{CB04} does converge. Thus, the grid method extends the iterative method for the ranges of $m$ where the iterative method does not converge.

Fig. \ref{fig:Left:i_crit_of_m} shows the numerical solution $m(i)$
such that $d\omega/d\tau(i,m(i))=0$. We see that for $m\ne0$, the
symmetry between the prograde and retrograde orbits breaks because
of the $\cos i$ terms. The red dashed line is the linear fit 
\begin{equation}
i(m)=i_{c,0}+i_{c,1}m,\label{eq:icrit2}
\end{equation}
where $i_{c,0}=39.2^{\circ}$(red point) and $i_{c,1}=71.3^{\circ}$
is calculated analytically in  appendix \ref{sec: app B}. 

Note that each $m$ is coupled to $\cos i$, thus the retrograde orbit
are accounted for with the convention that for retrograde orbit $m<0$
and $i\to\pi-i$. In this convention, $i$ is always smaller than $90^{\circ}$ (see \cite{CB04} for details and references therein). 

In order to use an explicit formula for $i(m)$, we have used a least
squares polynomial fit for the data of degree 40. The first few coefficients
are 
\begin{equation}
i(m)=39.36+67.47m-125.62m^{2}+\mathcal{O}(m^{3}).\label{eq:icrit3}
\end{equation}

The full list of the coefficients is given in table \ref{tab:2} in Appendix \ref{sec: app B}.
Note that the first two coefficients in Eq. (\ref{eq:icrit3}) are consistent
with the analytical solution in Eq. (\ref{eq:icrit2}). The right panel
of Fig. \ref{fig:Left:i_crit_of_m} shows the deviation of the polynomial fit from the numerical solution, with RMS value of  $\Delta i_{RMS}=0.08^{\circ}$. 

In our investigation of the stability limit, the orbits have relative
large separations, and thus large $m$ values. The transformation
from $a$ to $m$ is 
\begin{equation}
\frac{a}{r_{H}}=3^{1/3}m^{2/3}\label{eq:a_to_m}
\end{equation}

\begin{figure*}
\begin{center}
\includegraphics[width=0.45\textwidth]{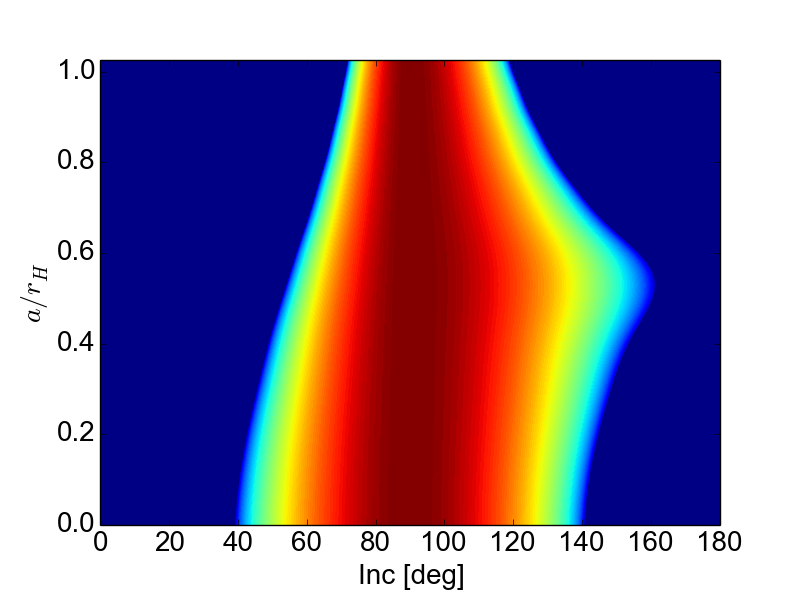}\includegraphics[width=0.45\textwidth]{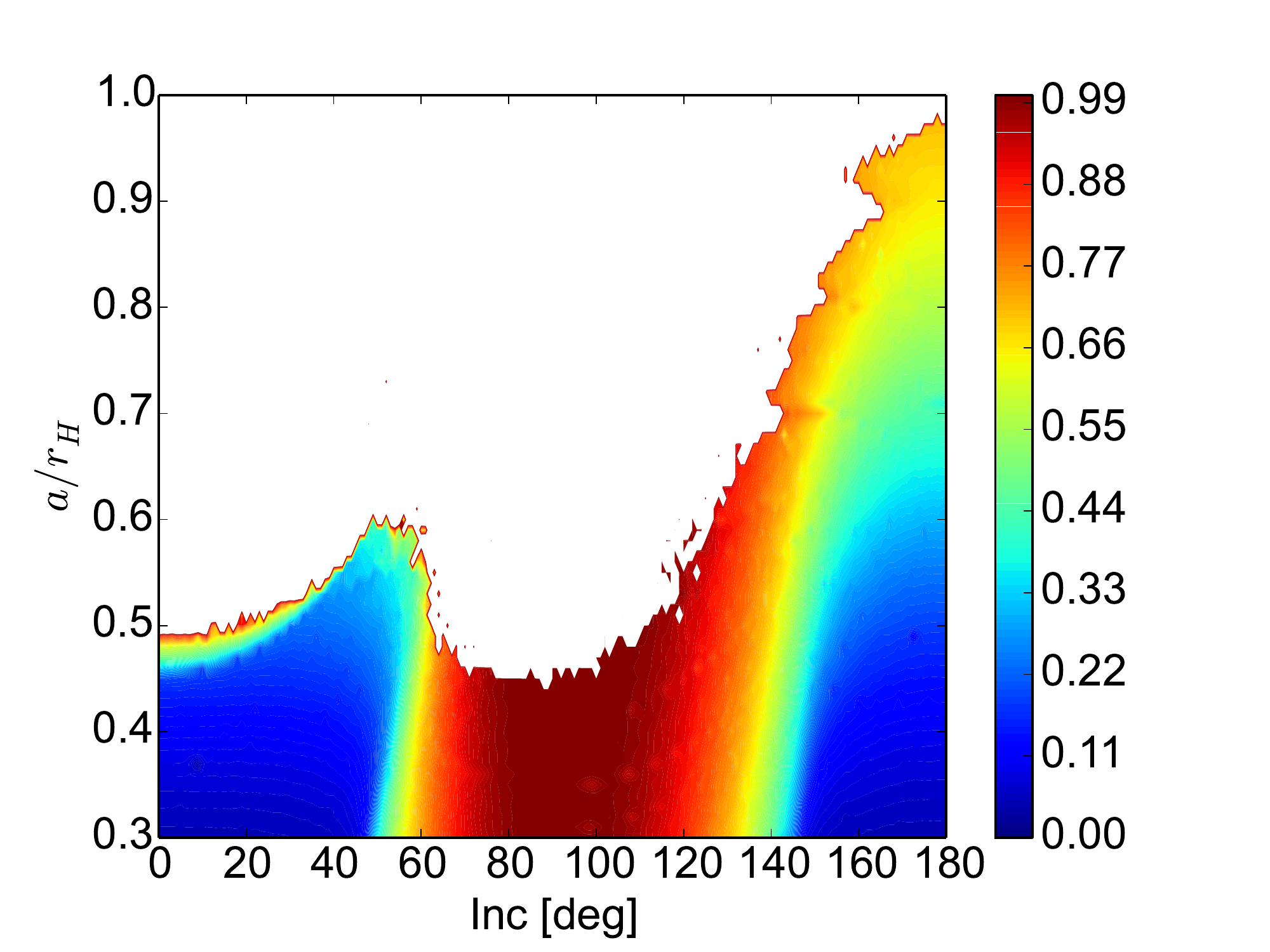}\caption{Left:  Maximal eccentricity of the
LK mechanism combined with ER and LT (Eq. \ref{eq:emax2}). Right: Simulated maximal eccentricity of the stable orbits. \label{fig:e_max}}
\end{center}
\end{figure*}

The maximal eccentricity is modified from Eq. (\ref{eq:emax}) to 
\begin{equation}
e_{max}(i,i_{c}(a))=\sqrt{1-\frac{\cos^{2}i}{\cos^{2}i_{c}(a)}}\label{eq:emax2}
\end{equation}
and is different for each $a$. In this case, the critical radii depends
on the maximal eccentricity (\ref{eq:emax2}) which in turn depends
on the critical radius. In order to overcome this technical difficulty
and find $i_{c}(a)$ in Eq. (\ref{eq:emax2}), we calculate first a
temporary value of $a$, 
$r_{temp}(a,i)\equiv r_{c}(i)\cdot f\left(e_{max}\left(a(m)\right)\right)$
 and look at $\Delta(a,i)\equiv r_{temp}(a,i)-a$. For small $a$,
$\Delta(a,i)>0$, or $a>r_{temp}(a,i)$ and the binary is stable.
Only when $\Delta(a,i)$ changes sign, $a\ge r_{temp}(a,i)$, and
the binary is unstable. Formally, the limiting radii is the minimum
of the set 
\begin{equation}
r_{c}(i)=\min_{a}\{\Delta(a,i)\le0\}.\label{eq:rcrti4}
\end{equation}

\begin{figure*}
\centering{}\includegraphics[width=0.45\textwidth]{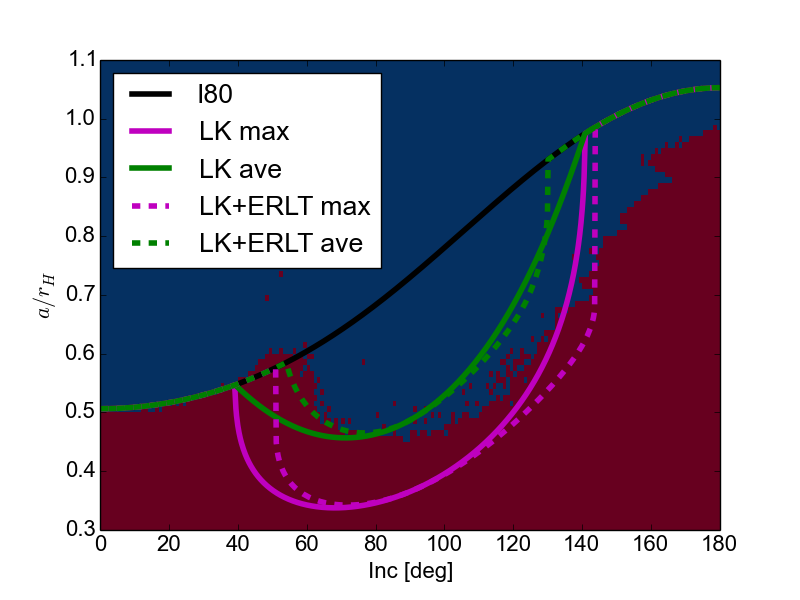}\includegraphics[width=0.45\textwidth] {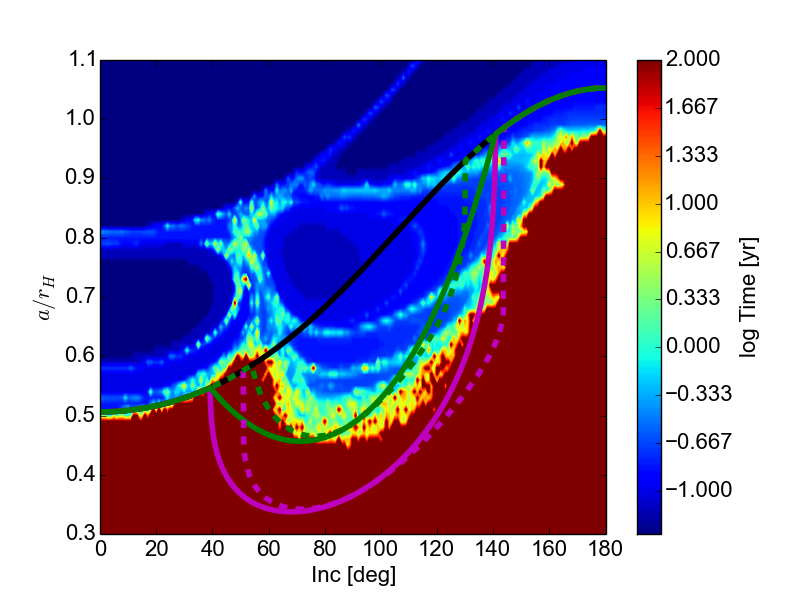} \caption{\label{fig:5}The same as Fig \ref{fig:2}, but
with additional fits where $e_{max}$ is determined both from LK and
ER-LT terms (Eq. \ref{eq:emax2}). The dashed green (magenta) lines
are fits due to both LK and ER-LT for $f_{ave}$ ($f_{max}$).
}
\end{figure*}

Fig. \ref{fig:e_max} shows the values of $e_{max}(i,a)$
given in Eq. (\ref{eq:emax2}). $e_{max}$ is restored for the standard
LK regime where $a\approx0$. As $a$ increases, the critical inclination
shifts to the right to higher values. Thus, for prograde orbits the
effective critical turn off point of the limiting radii curves will
be shifted to the right. For retrograde orbits the shift will occur
also to the right as long as $a/r_{H}\lesssim0.6$. For higher values,
the shift will be to the left. The right panel shows the resulted
maximal eccentricity map of the numerical integration. Overall, the numerical
results are in good agreement with the analytical expectations for
orbits with $a\lesssim0.6r_{H}$. For orbits with $a\gtrsim0.6r_{H}$
the perturbations in eccentricity are large. Thus, the orbits are
not simple Keplerian ellipses and the DA approximation breaks down.

We find the limiting radii in Eq. (\ref{eq:rcrti4}), and plot in Fig.
\ref{fig:e_max} the additional corrections to $r_{c}$ due
to the evection terms. Due to the effective shift in $i_{c}$ due
to evection, most of the bump near $\sim60^{\circ}$ is stable. The best fit model is the green dashed curve (LK and ERLT) is consistent with
the numerical stability map for inclinations up to $i\lesssim120^{\circ}$,
where in the retrograde cases it fails to explain the stability map. 

\section{Numerical parameter space exploration}
\label{sec:Numerical-parameter-space}

\subsection{Numerical set up}
\label{subsec:Numerical-set-up}

In order to test the critical stability radius, we have made numerical experiments, repeating, but also much extending the results of \cite{HB91}. The planetary orbit is
circular with semi-major axis $a_{out}=1$AU. The mass ratios are
$m_{1}/m_{out}=10^{-6}$ and $m_{2}/m_{1}=10^{-2}\ll 1$, thus the Hill
radius $r_{H}=a_{out}(10^{-6}/3)^{1/3}\ll a_{out}$. The test mass $m_2$ is non-zero, but it is small enough for having excellent corresponce between the numerical and analytic results. We simulate a
large grid of initial conditions of $a/r_{H}\in[0.3,1.1]$ with steps
$\delta a=0.01r_{H}$ and $i\in[0^{\circ},180^{\circ}]$ with steps
of $\delta i=1^{\circ}$. We start the runs with the line of ascending node $\Omega=0$, and
the mean anomaly $M=0$. The argument of periapse is ill
defined for circular orbits. The total number of simulations is $N=81\times181=14461$.
The longest LK timescale is $T_{LK}=\sqrt{3}P_{out}\cdot0.3^{-3/2}\approx10.5$
years, while the orbits near the stability limit have much shorter
timescale. 

We conclude that a specific orbit is unstable in two cases:
(i) If the binary eccentricity exceeds unity ($e>1)$. (ii) If the binary distance is larger than three times the Hill radius
$(d_{bin}>3r_{H}$). 
We use both criteria (i) and (ii) independently. Both criteria yield
the same stability maps.

\begin{figure*}
\begin{centering}
\includegraphics[width=0.45\textwidth]{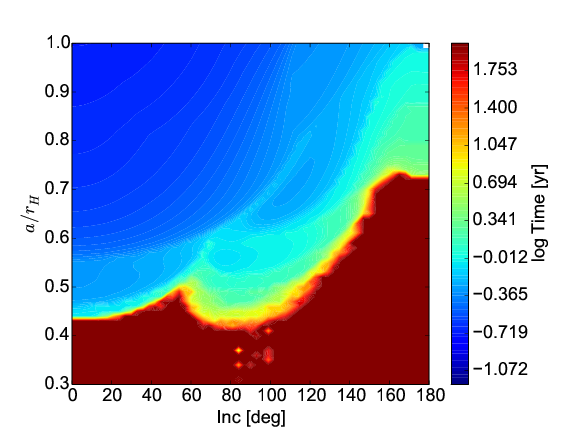}\includegraphics[width=0.45\textwidth]{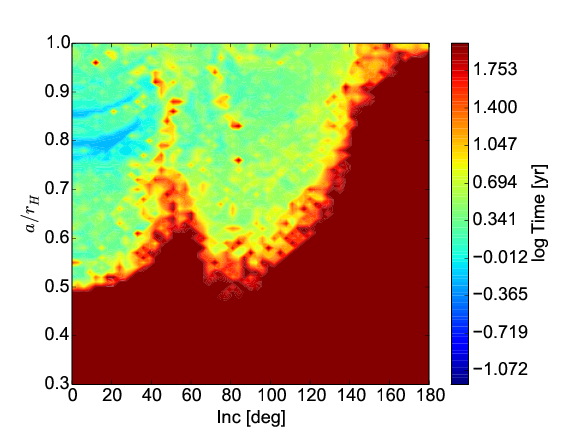}
\par\end{centering}
\begin{centering}
\includegraphics[width=0.45\textwidth]{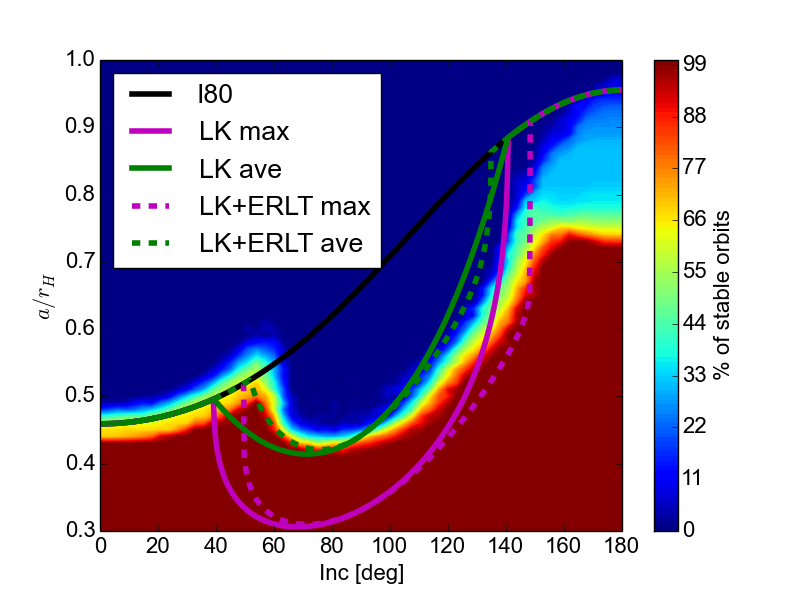}\includegraphics[width=0.45\textwidth]{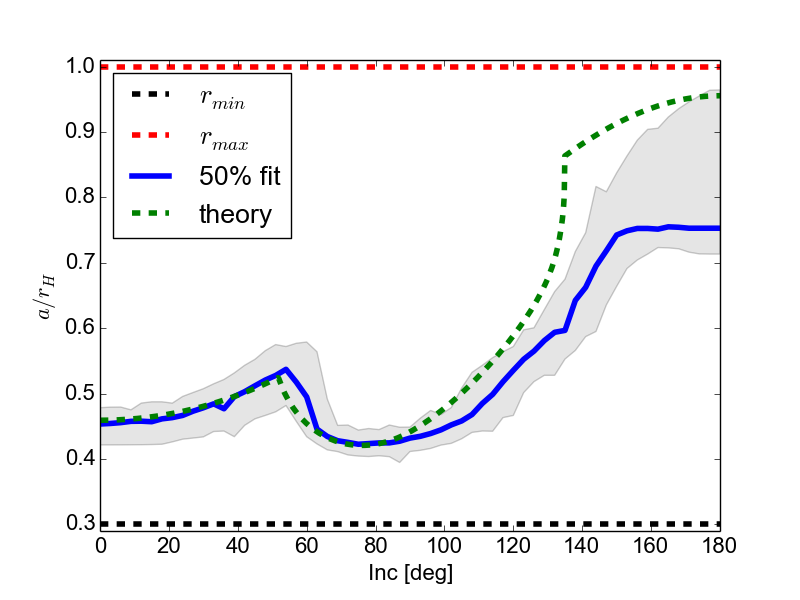}
\par\end{centering}
\caption{\label{fig:6} Stability map of the circular grid,
with $100$ different realizations of each $a$ and $i$. Top left:
Time map of the minimal time of break up for the $100$ realizations.
Top right: Time map of the maximal time of break up for the $100$
realizations. Bottom left: Percentage of the survived orbit with the
analytical fits. Note different normalization of the analytical fits.
Bottom right: Statistics of the destruction of the circular grid.
The blue line corresponds to the survival of $50\%$ of the orbits,
the green line corresponds to the best fit analytical model (KL+ERLT
ave). Grey region represents $95\%$ confidence levels.}
\end{figure*}

We simulate each initial condition for $100$ years, at least one
order of magnitude larger than the maximal LK timescale to capture
the (quasi) secular effects. We use an N-body integrator with a shared
but variable time step, using the Hermite 4th order integration scheme
following \citet{Hut95}.

The initial condition chosen to build the stability map of Fig. \ref{fig:5} 
limit the simulated parameter space to a degenerate subspace. The next sections explore more of the parameter space and
thus avoid any potential biases in the choice of the initial conditions.
We will show that the overall morphology of the stability map is conserved
and that the changes in the stability map are minor.

\subsection{Circular grid}
\label{numB}
In order to explore more of the parameter space, we repeat the simulation,
but now with a resolution of $\delta i=3^{\circ}$, and $a/r_H \in [0.3,1]$ with $\delta a = 0.01 r_H$. In addition,
we randomly draw $100$ initial conditions for the pair $(M,\Omega)$
for each value of $a$ and $i$. Since the inner binary is circular,
$\omega$ is not defined and not sampled. Thus, the total number of
initial conditions is $N=71\times61\times100=433,100$. 

\subsection{Eccentric grid}
\label{numC}
\begin{figure*}
\begin{centering}
\includegraphics[width=0.45\textwidth]{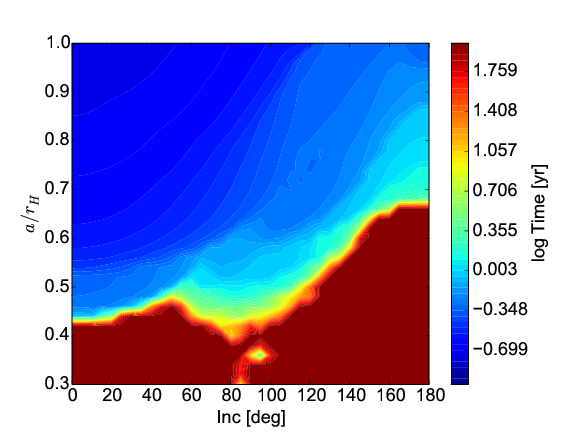}\includegraphics[width=0.45\textwidth]{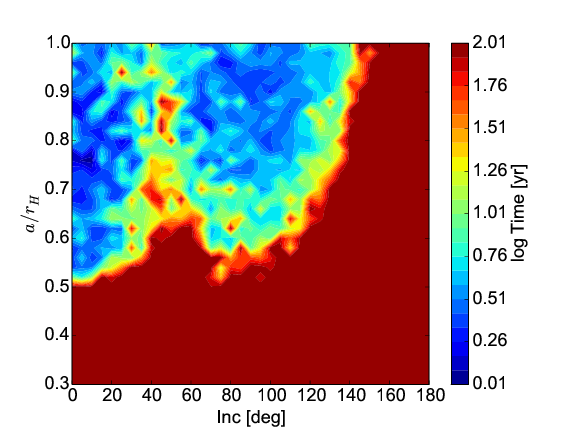}
\par\end{centering}
\begin{centering}
\includegraphics[width=0.45\textwidth]{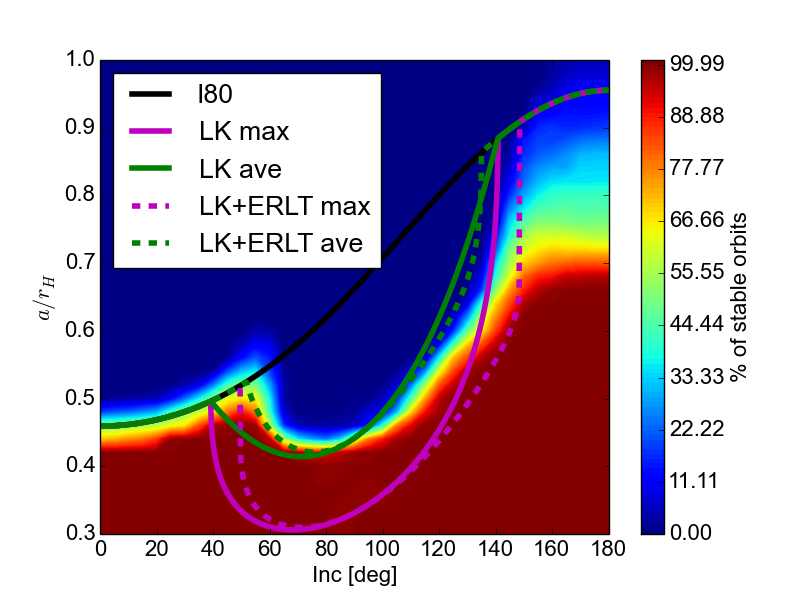}\includegraphics[width=0.45\textwidth]{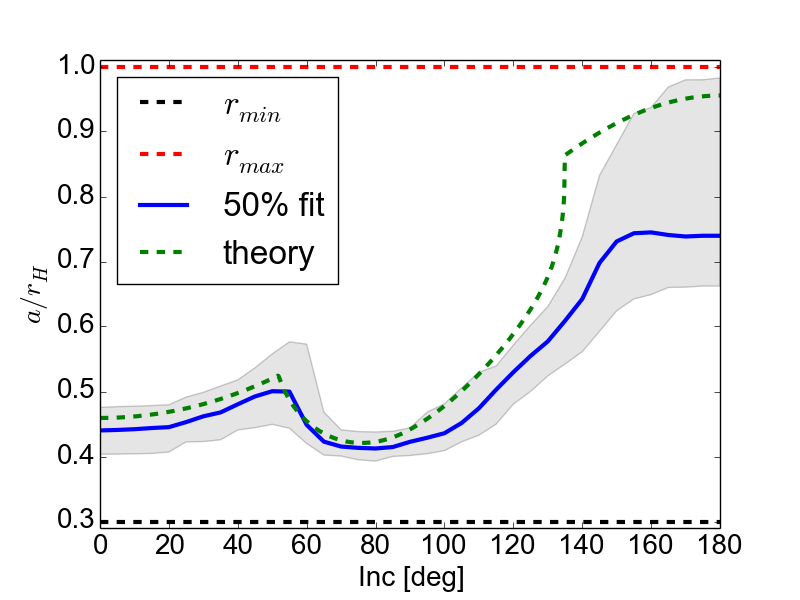}
\par\end{centering}
\caption{\label{fig:7}The same as Fig. \ref{fig:6},
but with initial eccentricity $e=0.1$ and sampling of the argument
of periapse $\omega$. }
\end{figure*}

Fig. \ref{fig:6} shows the stability map of the circular
grid. Top left (right) panel show the minimal (maximal) break up times
of the stability map. The bottom left panel shows the percentage of
the sampled stable orbits. We see that the stability limit has a chaotic
boundary, and some width of $\sim0.06r_{H}$. The orbits with zero
mean anomaly are the most stable ones. Thus, we normalized the analytical
fits by a fudge factor of $0.92$, such that they lie roughly in the
middle of the boundary for zero inclinations. Though the stability
limit is lower, the general morphology of the stability boundary is
preserved. For highly retrograde orbits, their stability limit has
a cut off at $\sim0.7-0.8r_{H},$ and some of the orbits above that
limit are unstable. The chaotic boundary is also wider. The bottom
right panel shows the comparison of the best analytical fit (green)
to the maximal radius with $50\%$ survival rate (blue). The shaded region is the $95\%$ confidence level; above
(below) the region, only $5\%$ or less of the orbits are unstable
(stable). We see that the analytical fit agrees well with the simulated
stability grid, up to radii of $\lesssim0.6r_{H}$ and inclination
of $\lesssim120^{\circ}.$ In addition, the $95\%$ confidence level
zone is wider for retrograde orbits, thus indicating that the $95\%$
confidence level is correlated to the chaotic boundary.

We repeat the simulations, but with $\delta i=5^{\circ}$, $\delta a = 0.02r_H$ and initial
eccentricity $e=0.1$. We randomly draw $343$ initial conditions for the
triple $(M,\Omega,\omega)$ for each value of $a$ and $i$. The,
the total number of initial conditions is $N=36\times37\times343=456,786$.

Fig. \ref{fig:7} shows the stability map of the
eccentric grid. Each panel is the same as in Fig \ref{fig:6}.
Besides slight changes in the width of the chaotic boundary and slightly
lower stability limit in the retrograde case ($a\lesssim0.7r_{H})$,
the results are very similar to the circular grid in Fig. \ref{fig:6}. 

In both the circular and eccentric cases, the analytical fit is better
since the angle of the ``turn-off point'' is slighlty lower than
$\sim60^{\circ}$. We conclude that the initial simulations are biased
toward the most stable orbits in the chaotic region, but still capture
the general morphological picture. The circular and eccentric grid
simulations help to overcome the bias and are a better fit for the
analytical theory.

\subsection{Polynomial fits}

\begin{table*}
\begin{centering}
\begin{tabular}{|c|c|c|c|c|c|c|c|c|}
\hline 
& $a_0$ & $a_1$ &$a_2$ &$a_3$ &$a_4$ &$q$ &$i_{c1}$ &$i_{c2}$ \tabularnewline
\hline 
$p_{th}$ & $7.211$ & $-14.9$ & $12.93$& $-5.05$& $0.75$& $0.721$ & $0.956 (54.8^{\circ})$ & $ 2.276	 (130.4^{\circ})$ \tabularnewline
\hline
$p_{circ}$ & $2.24$ & $-2.03$ & $0.6$& $0.68$& $-0.033$& $0.66$ & $0.94 (54^{\circ})$ & $ \pi (180^{\circ})$ \tabularnewline
\hline 
$p_{ecc}$ & $2.6$ & $-2.9$ & $1.3$& $-0.16$& $-0.0049$& $0.635$ & $0.96 (55^{\circ})$ & $ \pi (180^{\circ})$ \tabularnewline
\hline 
\end{tabular}
\par\end{centering}
\caption{\label{tab:1}Polynomial fit coefficients for the polynomials defined by Eq. (\ref{eq:pfit}). $p_{th}$ is the best analytic fit. $p_{circ}$ is the polynomial fit of the numerical circular grid (Fig. \ref{fig:6}). $p_{ecc}$ is the polynomial fit of the numerical eccentric grid (Fig. \ref{fig:7}). }
\end{table*}

For completeness, we provide 4th order polynomial fits for the best fit analytic model, the numerical results of the circular and eccentric grids. The polynomials are defined by

\begin{equation} 
r_{th}(i)/r_H=q(g(i))^{-2/3}\times\left\{ \begin{array}{cc}
p(i) & i_{c1}<i<i_{c2}\\
1 & \text{else}  
\end{array} \right.\label{eq:pfit} 
\end{equation} 

where $g(i)$ is given in Eq. (\ref{eq:rcrit2}), $q$ is the fudge factor and the polynomial $p(i)=\sum_{j=0}^{4} a_j i^j$ is the best 4th order fit for the data. Coefficients and parameters of the fitting polynomials  are summarized in Table \ref{tab:1}.

\section{Discussion and summary}
\label{sec:Discussion-and-summary}
In this paper we have developed a generalized stability criteria for triple
systems with arbitrary mutual inclination, and extended the Hill stability
criteria to account for secular evolution and evection resonance.
We used analytic arguments to derive the critical stability criteria
and complemented them with extensive three-body simulations. Using these
we also provided a convenient polynomial fitting formula for the stability criteria.
The comparison between the analytical theory and numerical integrations
is excellent, reproducing the morphology of the stability map, the
amplitude and the "turn-off point" of the bowl-like-shaped stability
region. In addition, the break up time-scales for highly inclined orbits
is comparable with the quasi-secular Lidov-Kozai timescale, and much
longer than the orbital period of the binary. This indicates that
the mechanism that drives the highly inclined binaries to instability
is indeed of secular nature, rather than a dynamical Hill instability.
In addition, the maximal eccentricity attained in the numerical integrations
is compatible to the maximal eccentricity predicted by the analytic theory.

Though the analytic theory well describes prograde orbits, there is
a significant discrepancy for retrograde orbits at high inclinations
beyond $i\gtrsim120^{\circ}$. At sufficiently large separations of
$a/r_{H}\gtrsim0.6$, the inner orbit of the triple is strongly perturbed
by the outer companion and the orbit is no longer Keplerian. In addition,
the timescales of the inner and outer binaries are comparable and
the averaging method we use is no longer valid. The latter is evident
in the eccentricity map of Fig. \ref{fig:e_max}, where
co-planar retrograde orbits, far from the Lidov-Kozai oscillation
regime, also evolve into large eccentricities. For retrograde inclinations
close to $180^{\circ}$, the analytic critical stability radius poorly
describes the critical stability radius. Properly sampling the parameter
space yields a chaotic transition boundary, which width increases
for retrograde orbits. In this regime the critical stability strongly
depends on the initial phase and orientation of the binary, such that
the transition from stable to unstable orbits stretches to the range
of $0.7\lesssim a/r_{H}\lesssim1$.

As an example we considered an application of our results for Solar
System satellites. Our findings shed light on the binomial inclination
distributions of irregular satellites \citep{2005AJ....129..518S,2015ARA&A..53..409W}.
In particular, the largest inclinations of prograde satellites is
near $\sim60^{\circ}$ ($57^{\circ}$ for Margaret, a moon of Uranus)
and $\sim130^{\circ}$ for retrograde satellites ( $\sim136.5^{\circ}$
for Neso, a moon of Neptune). At least for prograde satellites, the
critical angle of $\sim60^{\circ}$ is well explained by the evection
corrections to the LK mechanism, as we show (see also \citealp{CB04}).

\section*{Achnowledgements}

We would like to thank Scott Tremaine for useful discussions during
the 33rd winter school of theoretical physics in Jerusalem. We acknowledge
support from the Israel-US bi-national science foundation, BSF grant
number 2012384, European union career integration grant \textquotedblleft GRAND,\textquotedblright{}
the Minerva center for life under extreme planetary conditions and
the Israel science foundation excellence center I-CORE grant 1829/12. 

\appendix

\section{Analytical derivation of the minimal inclination}
\label{sec: app B}
The implicit Eq. (\ref{eq:icrit}), is satisfied at least at one
point $p_{c}=(i=i_{c,0},m=0)$. We assume that exists a local, unique
solution $i\equiv i(m)$ in the vicinity of the point $p_{c}$.

 Since $(d/di)h_{0}(i_{c})\ne0$, the assumptions of the implicit function
theorem are satisfied, and a solution $i(m)$ can be found locally
as a power series in $m$. Deriving with respect to $m$ Eq.( \ref{eq:icrit})
yields the first derivative: 
\begin{equation}
i_{c,1}=\left.\frac{di(m)}{dm}\right|_{m=0}=\left.-\frac{h_{1}(i)\cos i}{h_{0}^{\prime}(i)}\right|_{i=i_{c,0}}\label{eq: implicit}
\end{equation}
 where $h_{0}^{\prime}(i)=(d/di)h_{0}(i)$. Substitution at $i=i_{c}$
and $m=0$ yields $i_{1}=71.3^{\circ}$.

\section{Coefficients of Lunar Theory}
\label{sec: app A}
We use the coefficients of Andoyer moritz, taken from table III of \citep{1971AJ.....76..273D}. These coefficients are highly accurate up to order $m^{10}$. \cite{CB04} neglect any high order term variation in $\dot{\Omega}$ (Note their Eqns. (35) and (36)) where the evolution of $\dot{\Omega}$ due to Lunar theory is absent. We include this evolution given in Eq. (A4) of \cite{SahaTremaine93} such  that $\dot{\omega}$ is extracted from $\dot{\Omega}$ and $\dot{\varpi}$. We use the coefficients $C_k = c_k(\varpi) - c_k(\Omega)$, which are given in table \ref{tab:c}. 

{\renewcommand{\arraystretch}{2}
\begin{table}
\begin{centering}
\begin{tabular}{|c|c|c|}
\hline 
& $c_k(\varpi)$ & $c_k(\Omega)$ \tabularnewline
\hline 

$3$ & $ \frac{225}{32}$ & $\frac{9}{32}$  \tabularnewline

\hline

$4$ & $\frac{4071}{128}$ & $\frac{273}{128}$ \tabularnewline

\hline 
$5$ & $\frac{265493}{2048}$ & $\frac{9797}{2048}$  \tabularnewline

\hline
$6$ & $\frac{12822631}{24576}$ & $\frac{199273}{24576}$ \tabularnewline
\hline
$7$ & $\frac{1273925965}{589824}$ & $\frac{6657733}{589824}$  \tabularnewline
\hline
$8$ & $\frac{66702631253}{7077088}$ & $-$ \tabularnewline
\hline 
$9$ & $\frac{29726828924189}{679477248}$ & $-$  \tabularnewline
\hline
$10$ & $214939.72297513$ & $-$ \tabularnewline
\hline
$11$ & $1102182.3514701$ & $-$ \tabularnewline
\hline 
\end{tabular}
\par\end{centering}
\caption{\label{tab:c} The Coefficients used in Eq. (\ref{eq:icrit}). Taken from \protect\cite{1971AJ.....76..273D} and \protect\cite{SahaTremaine93} }
\end{table}

\section{Coefficients of the least squares polynomial}
\label{sec: app C}

{\renewcommand{\arraystretch}{1}
\begin{table}
\begin{centering}
\begin{tabular}{|c|c|c|c|}
\hline 
1-10 & 11-20 & 21-30 & 31-40\tabularnewline
\hline 
\hline 
$67.47$ & $1.2\cdot10^{10}$ & $-1.93\cdot10^{14}$ & $-6.81\cdot10^{15}$\tabularnewline
\hline 
$-125.62$ & $-5.64\cdot10^{9}$ & $6.85\cdot10^{13}$ & $-1.95\cdot10^{15}$\tabularnewline
\hline 
$2122.95$ & $-1.61\cdot10^{10}$ & $4.54\cdot10^{14}$ & $9.77\cdot10^{15}$\tabularnewline
\hline 
$11334.5$ & $1.03\cdot10^{11}$ & $-7.16\cdot10^{13}$ & $5.8\cdot10^{15}$\tabularnewline
\hline 
$-269265.4$ & $1.52\cdot10^{12}$& $-6.58\cdot10^{14}$ & $-8.57\cdot10^{15}$ \tabularnewline
\hline 
$-674029.4$ & $-1.03\cdot10^{12}$ &  $-1.378\cdot10^{14}$ & $-7.56\cdot10^{15}$ \tabularnewline
\hline 
$1.72\cdot10^{7}$& $-1.03\cdot10^{13}$
& -$6.2\cdot10^{13}$ & $4.34\cdot10^{15}$ \tabularnewline
\hline 
$1.32\cdot10^7$ & $6.53\cdot10^{12}$& $5.61\cdot10^{14}$ & $5.08\cdot10^{15}$\tabularnewline
\hline 
$-5.85\cdot10^8$& $5.1\cdot10^{13}$
& $2.6\cdot10^{15}$ & $-9.76\cdot10^{14}$ \tabularnewline
\hline 
$4.51\cdot10^{8}$ & $-2.7\cdot10^{13}$
&$-3.41\cdot10^{14}$ & $-1.43\cdot10^{15}$ \tabularnewline
\hline 
\end{tabular}
\par\end{centering}
\caption{\label{tab:2}First 40 coefficients of the polynomial defined in Eq.  (\ref{eq:icrit3}).}
\end{table}

\bibliographystyle{mnras}


\end{document}